\documentclass[aps,preprint,showpacs]{revtex4}
\usepackage{amsmath}
\newcommand{\be}{\begin{equation}}
\newcommand{\ee}{\end{equation}}
\newcommand{\bea}{\begin{eqnarray}}
\newcommand{\eea}{\end{eqnarray}} 
\newcommand{\ba}{\begin{array}}
\newcommand{\ea}{\end{array}}

\newcommand{\bb}{\bibitem}

\begin{document}

\title{\bf Susceptibility amplitude ratio for generic competing systems}
\author{C. F. Farias\footnote{e-mail:cffarias@fisica.ufpb.br}}
\affiliation{{\it Departamento de F\'\i sica, Universidade Federal da 
Para\'\i ba, Caixa Postal 5008, 58051-970, Jo\~ao Pessoa, PB, Brazil}} 
\author{Marcelo M. Leite\footnote{e-mail:mleite@df.ufpe.br}}
\affiliation{{\it Laborat\'orio de F\'\i sica Te\'orica e Computacional, Departamento de F\'\i sica,\\ Universidade Federal de Pernambuco,\\
50670-901, Recife, PE, Brazil}}

%\end{center}
\vspace{0.2cm}
\begin{abstract}
{\it We calculate the susceptibility amplitude ratio near a generic higher 
character Lifshitz point up to one-loop order. We employ a renormalization 
group treatment with $L$ independent scaling transformations associated to the 
various inequivalent subspaces in the anisotropic case in order to compute 
the ratio above and below the critical temperature and demonstrate 
its universality. Furthermore, the isotropic results with only one type of 
competition axes have also been shown to be universal. We describe how the 
simpler situations of $m$-axial Lifshitz points as well as ordinary 
(noncompeting) systems can be retrieved from the present framework.}   
\end{abstract}

\vspace{1cm}

\maketitle

\newpage
\section{Introduction}
\par The occurrence of $m$-axial Lifshitz phase transitions \cite{Ho-Lu-Sh,Ho} 
in various real physical systems (e.g., magnetic modulated materials 
\cite{Be,Yo1,Yo2,Be1,Hana,We}, high-$T_{c}$ superconductors \cite{H,K,S}, 
liquid crystals \cite{Ra,Za,Ska}, etc.) has increased the interest in the 
field-theoretic description of this subject 
in the last few years \cite{MC,AL,DS}. According to modern renormalization 
group arguments, critical phenomena of $m$-axial Lifshitz competing systems
have their universality classes characterized by $(N,d,m)$, namely, the number 
of components of the (field) order parameter $N$ dwelling in $d$ space 
dimensions with $m(\equiv m_{2})$ space directions presenting alternate 
(repulsion-attraction) among the fields \cite{L,Leite1,CL1}. Another type of 
``competing axes'' can be defined: if the alternate couplings are of the type 
attractive-repulsive-attractive and take place along $m_{3}$ spatial 
dimensions, one refers to $m_{3}$-axial third-character Lifshitz critical 
behavior (for the realization of the $m_{3}=1$ case, see \cite{Se1}). 
\par More generally, 
a $m_{L}$-fold $L$-th character Lifshitz behavior appears whenever 
short-ranged alternate interactions with L couplings of the type 
repulsion-attraction-repulsion-attraction-... are allowed \cite{Se2,Ni1,Ni2}.
If all sorts of the aforementioned competing axes are present simultaneously 
in the critical system under consideration, its phase transitions are 
governed by the generic higher character Lifshitz critical behavior. The 
universality classes of these arbitrary competing systems are defined by the 
set $(N,d,m_{2},...,m_{L})$ \cite{Leite2,LeiteII,CL2}. The language of 
magnetic systems is particularly suitable to describe these systems. It 
is convenient to make the connection of these complex critical behaviors 
with the prototype of second order phase transitions in noncompeting systems: 
the Ising model ($N=1$) \cite{Amit}.
\par The simplest realization of the usual Lifshitz universality class 
$(N,d,m)$ can be encountered in uniaxial critical (systems $m=1$) behavior. 
It can be understood in terms of the 
axial next-nearest-neighbor Ising (ANNNI) model \cite{Selke1,Selke2} which 
corresponds to the usual Ising model including antiferromagnetic exchange 
interactions among second neighbor spins along a single axis in a cubic 
lattice. The uniaxial Lifshitz point arises at the confluence of the 
disordered, a uniformly ordered and a modulated phase. Anisotropic second 
character $m$-axial points generalize that uniaxial when the 
competing axes occur along $m$ space directions whenever $m \neq d$. 
\par The ANNNI model can be generalized by including further ferromagnetic 
couplings among third neighbors along the competing axis. Except for a 
little additional complication in the phase diagrams due to the existence of 
an additional parameter related with the third neighbor coupling, similar 
phases can be defined such that the region of intersection of them 
terminates in a point where the different phases characterizing the system 
meet, known as the uniaxial Lifshitz point of third character \cite{Se1}. 
This reasoning can be extended to contemplate the situation where 
alternate couplings up to the $L$-th neighbors exist, and the critical 
point associated to the region of confluence of the several phases of the 
system is denominated the uniaxial Lifshitz point of $L$-th character. If the 
system presents competing interactions of this type along $m_{L}\neq d \;(=d)$ 
space directions, the system is said to have anisotropic (isotropic) 
critical behavior with $m_{L}$-axial point of $L$-th 
character \cite{Se2,Ni1,Ni2}. The modulated phases have a distinction when we 
compare anisotropic and isotropic behaviors. In the former there exist two 
types of correlation lengths, namely $\xi_{1}$ and $\xi_{L}$, which label two 
inequivalent subspaces characterized by correlations perpendicular and 
parallel, respectively, to the $m_{L}$ subspace. In isotropic behaviors  
only one type of correlation length $\xi_{L}$ characterizes each modulated 
phase.
\par The anisotropic Lifshitz point of generic $L$th character can also be 
defined in the description of the most general $d$-dimensional competing
system, whenever several types of competing axes show up {\it simultaneously}. 
Let us consider its simplest realization. In that case, there are 
only nearest neighbor interactions along $m_{1} \equiv (d-m_{2} -...-m_{L})$ 
(noncompeting) directions,
second neighbor competing interactions along $m_{2}$ directions perpendicular 
to the $m_{1}$ dimensions, competition among third neighbors along 
$m_{3}$ space directions (orthogonal to the $(m_{1},m_{2})$ subspaces), etc., 
up to $L$th neighbor alternate couplings along $m_{L}$ directions, with all 
competition subspaces perpendicular to each other. The 
model which describes this sort of arbitrary competing systems was proposed 
a few years ago and named {\it competing exchange coupling Ising} 
($CECI$) model \cite{Leite2}. There are $L$ 
inequivalent correlation lengths owing to the $L$ independent competing axes 
$m_{n}$ ($n=1,..,L$). This situation allows in principle low temperature 
$(L-1)$ modulated phases in equilibrium with the uniform ordered phase as 
well as with the high temperature disordered phase close to the Lifshitz point.
In addition, it is also possible that the complex systems can display 
several low temperature (up to $L-1$) uniformly ordered phases in equilibrium 
with (at least one) modulated ordered and high temperature disordered phases. 
\par The isotropic $m$-axial critical behavior has been 
experimentally realized in the context of polymers. At first the isotropic 
behavior was thought of being of purely academic 
interest. Nevertheless, its theoretical mean-field prediction in 
copolymer-homopolymer ternary blends \cite{BF} and 
subsequent experimental identification in mixtures of 
block copolymer-homopolymer \cite{Bates1} caused a certain enthusiasm, but the 
following paper with a more detailed analysis on the subject showed a 
microemulsion phase incompatible with the existence of the Lifshitz point 
\cite{Bates2}. It was argued there that the fluctuations destroyed this
multicritical point, although the associated critical region could be 
identified with the vicinity of the would be mean field Lifshitz point. The theoretical effect 
of fluctuations was incorporated 
immediately afterward using a self-consistent field theory (SCFT), and the 
ordered lamellar phase previously identified as the modulated phase 
was understood 
utilizing a one-component order parameter ($N=1$) \cite{KM}. This result not 
only confirmed the previous discussion from Ref.\cite{Bates2}, but also 
located loci in the mean field phase diagram with third character isotropic 
Lifshitz point. Later, mean field studies using SCFT indicated the existence 
of up to $6$th character isotropic critical behavior in blends of 
diblock copolymers \cite{OH}. This suggests that the isotropic behaviors of 
the $CECI$ model might be useful in unveiling properties of these real 
physical systems, even though the anisotropic realization of this model has 
not been identified yet.
\par Thence, several experiments have been performed for these polymers. 
For instance, the susceptibility of a homopolymer-diblock copolymer blend 
(polybutadiene and polystyrene) has been investigated recently using small 
angle neutron scattering. Some amplitudes above the Lifshitz temperature 
were estimated for fixed values of the diblock copolymer composition 
\cite{Pipi}. The closest we can get to this system using field theory 
techniques is to look for universal quantities related to the susceptibility, 
i.e., the amplitude ratio above and below the Lifshitz critical temperature 
\cite{Leite3}. By the same token, the study of a 
simple property such as the susceptibility for the $CECI$ model could shed 
light on possible future experiments related with real physical systems 
manifesting this especial critical behavior. 
\par In this paper, the 
susceptibility amplitude ratio for generic competing systems will be computed 
using field theory and renormalization group arguments up to one-loop 
order in a perturbation expansion. The anisotropic behaviors with arbitrary 
types of competing axes are discussed first. We are 
going to restrict ourselves to (fields) order parameters of only one 
component ($N=1$). The results are presented in a manifestly universal form 
and are shown how to reduce to the ordinary amplitude ratio without 
competition. We then restrict the number of 
competition axes to obtaining information on particular universality classes, 
the most obvious being the $m$-axial anisotropic Lifshitz criticalities. We 
show that the uniaxial result can be retrieved from the arbitrary anisotropic 
competing systems in a simple manner. The 
isotropic amplitude ratios for isotropic critical behaviors are calculated for 
the first time. The $m$-axial universality class is recognized from the 
generic situation whenever $n=2$. We show that although the expansion 
parameter is large for three-dimensional systems, our perturbative results are 
meaningful for those systems. As an application, we compare our 
field-theoretic $m=3$ isotropic output with experimental results from 
homopolymer-diblock copolymer mixtures and show very good agreement among 
them. 
\par The paper is organized as follows. In Section II we present the one-loop 
effective potential. We highlight a brief explanation of the several subspaces 
which occur in the problem along with the independent renormalization group 
transformations in the anisotropic cases. A simpler analogous discussion 
for calculating the isotropic amplitude ratio is explicited in Section III. 
Section IV presents the discussion of the results and conclusions.    

\section{Anisotropic amplitude ratio for generic competing systems}
\par We begin with the bare Lagrangian 
density associated with anisotropic generic competing systems described by 
the $CECI$ model, which is given by
\begin{eqnarray}\label{1}
L &=& \frac{1}{2}
|\bigtriangledown_{(d- \overset{L}{\underset{n=2}{\sum}} m_{n})} \phi_0\,|^{2} +
\sum_{n=2}^{L} \frac{\sigma_{n}}{2}
|\bigtriangledown_{m_{n}}^{n} \phi_0\,|^{2} \\ \nonumber
&& + \sum_{n=2}^{L} \delta_{0n}  \frac{1}{2}
|\bigtriangledown_{m_{n}} \phi_0\,|^{2}
+ \sum_{n=3}^{L-1} \sum_{n'=2}^{n-1}\frac{1}{2} \tau_{nn'}
|\bigtriangledown_{m_{n}}^{n'} \phi_0\,|^{2} \\ \nonumber
&&+ \frac{1}{2} t_{0}\phi_0^{2} + \frac{1}{4!}\lambda_0\phi_0^{4} .
\end{eqnarray}
\par The parameters which correspond to the physical situations are the 
coefficients of the derivatives of the bare field $\phi_0$ (order parameter of 
the phase transition), the bare reduced temperature 
$t_{0} (\propto T-T_{L}$, where $T_{L}$ is the Lifshitz critical temperature) 
and the bare coupling constant $\lambda_0$.
\par The Lifshitz critical region is defined for particular combinations of 
the exchange interactions among all the neighbors. This implies the
fine-tuning conditions on some parameters, namely, 
$\delta_{0n} = \tau_{n n'} =0$. 
\par At the Lifshitz critical region, the 
temperature is close but not equal to $T_{L}$. The structure of the field 
theory considered at this region is such that its momentum dependence on the 
various competing subspaces is rather peculiar. There are 
quadratic momenta components along the $m_{1}$-dimensional 
noncompeting subspace, quartic momenta components along the 
$m_{2}$-dimensional competing subspace, and so on, up to the $2L$-th power 
of momenta along the $m_{L}$-dimensional subspace, which are present 
simultaneously in the free bare critical propagator in momentum space. We can 
set $\sigma_{n}=1$ provided we perform a dimensional redefinition in the 
momentum characterizing the $n$-th $m_{n}$-dimensional competition subspace. 
If $\Lambda$ is a momentum scale, we take the engineering dimension of the 
competing subspace as $[k_{(n)}]=\Lambda^{\frac{1}{n}}$.   
\par The anisotropic behaviors is characterized by $L$ independent correlation 
lengths $\xi_{n}$, one for each subspace. They induce $L$ independent 
renormalization group flows in the parameter space of the massless 
theory. If we use normalization conditions in the definition 
of the renormalized theory, these flows can be described by $L$ independent 
sets of normalization conditions, each of them defining a symmetry point 
$SP_{n}$ ($n=1,..., L$) which simplifies our task of computing universal 
quantities in this formalism of one-particle irreducible ($1PI$) vertex 
parts. 
\par Typically, The Feynman integrals involved depend on
various external momenta scales, namely that characterizing the 
$(d-m_{2}-...-m_{L})$-dimensional noncompeting subspace, a momentum scale 
associated to the $m_{2}$ space directions, etc., up to the momentum scale 
corresponding to the $m_{L}$ competing axes. For example, an explicit integral 
that shall be used is the one-loop contribution to the four-point function, 
namely
\begin{equation}\label{2}
I(P,K'_{(2)},...,K'_{(L)}) =  \int \frac{d^{(d-\overset{L}{\underset{n=2}{\sum}} m_{n})}q \overset{L}{\underset{n=2}{\prod}} d^{m_{n}}k_{(n)}}
{[\overset{L}{\underset{n=2}{\sum}}\bigl((k_{(n)} + K_{(n)}^{'})^{2}\bigr)^{n} +
(q + P)^{2}] \left(\overset{L}{\underset{n=2}{\sum}}(k_{(n)}^{2})^{n} + q^{2}  \right)}\;\;\;.
\end{equation}
Although this integral should be computed in arbitrary 
nonvanishing external momenta 
components, in practice the calculation is simplified when we choose only one 
subspace, say $m_{n}$, whose momenta are set in the arbitrary value 
$K^{' 2n}_{(n)}= \kappa_{n}^{2n}$. In case we wish to determine universal 
quantities associated to vertex parts 
along the $j$th type of competing axes, we set $\kappa_{n}=0$ 
for $n \neq j$ maintaining, however, $\kappa_{j} \neq 0$. 
\par Unfortunately, the 
integral cannot be solved exactly for arbitrary 
external momenta, but can be resolved using the orthogonal approximation which 
permits to obtain the integral as an homogeneous function of the external 
momenta. Within this framework, the Feynman integrals of the corresponding 
field theory can be computed to all loop orders. In normalization conditions, 
the results for those integrals are independent of the subspace chosen. 
\par In order to be precise in our description, we should label the 
renormalized vertex parts according to the subspace characterized by the 
nonvanishing momenta scale $\kappa_{n}$ associated to the symmetry point 
$SP_{n}$. Fortunately, we do not have to employ this label in the present 
work, since in the context of the orthogonal approximation all renormalization 
directions possess the same fixed point \cite{Leite2}. In other words, the 
susceptibility amplitude ratio is independent of the renormalization group 
transformation characterized by the 
variation of the external momenta scale $\kappa_{n}$ in the renormalized 
vertex parts. In what follows, we shall use for simplicity all vertex parts 
computed at the symmetry point $SP_{1}\equiv SP$, i.e., the external momenta 
scale is given by $\kappa_{n}=\delta_{1n}\kappa$ with $P^{2}=\kappa^{2}=1$. 
With this choice we do not need to employ the orthogonal approximation 
to perform this integral in the anisotropic case, since the nonvanishing 
external momenta is contained in the quadratic term of the propagator. This 
quadratic part can be evaluated to arbitrary external momenta and the 
resulting expression for this integral is exact. For further details, the 
reader is advised to consult Ref. \cite{LeiteII}. 
\par The bare quantities can be transformed into renormalized amounts at 
one-loop level through the renormalization of the bare field and 
temperature which are given by $t_{0} = Z_{\phi^{2}}^{-1}t,
\phi = Z_{\phi}^{-\frac{1}{2}}M$ beside the renormalized coupling 
constant. We express the latter in the fixed point in terms of the 
dimensionless entity $u^{*}$ as $g^{*}=u^{*} \kappa^{\epsilon_{L}}$ where 
$\epsilon_{L} = 4 + \overset{L}{\underset{n=2}{\sum}}\frac{(n-1)}{n} m_{n} - d$ is the 
perturbation parameter. 
\par Using the symmetry point, let us write down the 
one-loop renormalized Helmholtz free energy density at the fixed point. 
It is simply the renormalized effective potential 
at one-loop plus polynomial terms in $t$ used to define additively 
renormalized vertex parts. For completeness (and anticipating future 
discussions for the specific heat amplitude ratio as well) we include a term 
proportional to $t^{2}$ which, however, will have no consequence to our 
discussion in the present work. Putting those arguments together, we obtain 
the following expression 
\begin{eqnarray}\label{3}
F(t,M)\;\; =&& \frac {1}{2}t M^{2} + \frac{1}{4!} g^{*} M^{4} +
\frac{1}{4}(t^{2} + g^{*} t M^{2} + \frac{1}{4} (g^{*} M^{2})^{2})
I_{SP} \nonumber\\
&& + \frac{1}{2} \int d^{(d-\overset{L}{\underset{n=2}{\sum}} m_{n})}q \Bigl[\overset{L}{\underset{n=2}{\prod}} d^{m_{n}}k_{(n)}\Bigr]
\Bigl[ln\Bigl(1 + \frac{t+ \frac{1}{2} g^{*} M^{2}}{\left(\overset{L}{\underset{n=2}{\sum}} (k_{(n)}^{2})^{n} + q^{2} \right)}\Bigr) \nonumber\\
&&- \frac{g^{*} M^{2}}{2\left(\overset{L}{\underset{n=2}{\sum}} (k_{(n)}^{2})^{n} + q^{2} \right)}\Bigr]\;\;,
\end{eqnarray}
where in the above equation $t, M$ ($t_{0} = Z_{\phi^{2}}^{-1}t,
\phi = Z_{\phi}^{-\frac{1}{2}}M$) are the renormalized (bare) reduced
temperature and order parameter, respectively, $Z_{\phi^{2}},
Z_{\phi}$ are normalization functions, $g^{*}$ is the renormalized
coupling constant at the fixed point, $\vec{q}$ is a
$(d-m)$-dimensional wave vector perpendicular to the competing axes,
whereas $\vec{k}$ is a $m$-dimensional wave vector whose components
are parallel to the competition axes. The integral $I_{SP}$ is defined
by:
\begin{equation}\label{4}
I_{SP} =  \int \frac{d^{(d-\overset{L}{\underset{n=2}{\sum}} m_{n})}q \overset{L}{\underset{n=2}{\prod}} d^{m_{n}}k_{(n)}}
{[\overset{L}{\underset{n=2}{\sum}} k_{(n)}^{2n} +
(q + P)^{2}] \left(\overset{L}{\underset{n=2}{\sum}} k_{(n)}^{2n} + q^{2}  \right)}\;\;\;,
\end{equation}
where the convenient symmetry point for this integral is defined as above, 
namely, $P^{2}= \kappa^{2}=1$. This choice has the virtue of transforming the 
dimensionful coupling constant in its dimensionless version, i. e., 
$g^{*} = u^{*}$ and is the most effective route to computing universal 
quantities in the context of the renormalization group strategy. Whenever a 
loop integral is performed, a typical geometric angular factor is produced, 
which can be factored out in a redefinition of the coupling constant in a 
standard way \cite{Amit}. In our case this factor is given by the expression 
$[S_{(d-\overset{L}{\underset{n=2}{\sum}} m_{n})} 
\Gamma(2 - \overset{L}{\underset{n=2}{\sum}}\frac{m_{n}}{2n})(\overset{L}{\underset{n=2}{\prod}} 
\frac{S_{m_{n}} \Gamma(\frac{m_{n}}{2n})}{2n})]$, such that it is 
going to be omitted whenever we report the result of any loop integral. 
The last integral at the symmetry point was already computed in 
Ref.\cite{LeiteII} and shown to be given by
$I_{SP} = \frac{1}{\epsilon_{L}}(1 + h_{m_{L}}\epsilon_{L})$, where 
$h_{m_{L}}= 1 + \frac{(\psi(1) - \psi(2- \overset{L}{\underset{n=2}{\sum}}\frac{m_{n}}{2n}))}
{2}$ and $\psi(z) = \frac{dln\Gamma(z)}{dz}$. It is worthy to stress that 
whenever $m_{3} = ...=m_{L}=0$, $h_{m_{2}}=[i_{2}]_{m}$ and the 
usual anisotropic $m$-axial Lifshitz critical behavior is obtained from this 
more general competing situation in a rather simple manner.
\par Since we need the value of $M$ in the coexistence curve above and below 
$T_{L}$, let us compute the renormalized magnetic field, which is given by
\begin{equation}\label{5}
H_{R}=\frac{\partial F}{\partial M}= tM + \frac{1}{6} u^{*}M^{3} 
+ \frac{u^{*}M}{2}(t+ \frac{u^{*}M^{2}}{2})[I_{SP}-I],
\end{equation}
where 
\begin{equation}\label{6}
I =  \int \frac{d^{(d-\overset{L}{\underset{n=2}{\sum}} m_{n})}q \overset{L}{\underset{n=2}{\prod}} d^{m_{n}}k_{(n)}}
{[\overset{L}{\underset{n=2}{\sum}}\bigl((k_{(n)})^{2}\bigr)^{n} + q^{2}] \left(\overset{L}{\underset{n=2}{\sum}}(k_{(n)}^{2})^{n} + q^{2} + t + \frac{u^{*}M^{2}}{2} \right)}\;\;\;.
\end{equation}
Let us compute explicitly this integral. First we use a Feynman parameter to 
write it as
\begin{equation}\label{7}
I =  \int_{0}^{1} dx \int \frac{d^{(d-\overset{L}{\underset{n=2}{\sum}} m_{n})}q \overset{L}{\underset{n=2}{\prod}} d^{m_{n}}k_{(n)}}
{[\overset{L}{\underset{n=2}{\sum}}\bigl((k_{(n)})^{2}\bigr)^{n} + q^{2} 
+ x(t + \frac{u^{*}M^{2}}{2})]^{2}}\;\;\;.
\end{equation}
In order to integrate the quadratic momentum out, we employ the identity
\begin{equation}\label{8} 
\int d^{d}q \frac{1}{[q^{2} + 2kq + m^{2}]^{\alpha}}= \frac{S_{d} 
\Gamma(\frac{d}{2}) \Gamma(\alpha - \frac{d}{2})}{2\Gamma(\alpha)} 
(m^{2}-k^{2})^{\frac{d}{2} - \alpha},
\end{equation}
and get to 
\begin{eqnarray}\label{9}
&& I = \frac{1}{2} S_{(d-\overset{L}{\underset{n=2}{\sum}} m_{n})} 
\Gamma(d - \overset{L}{\underset{n=2}{\sum}}m_{n}) \Gamma(2 -(d - \overset{L}{\underset{n=2}{\sum}}m_{n})) 
\int_{0}^{1} dx \int \overset{L}{\underset{n=2}{\prod}} d^{m_{n}}k_{(n)}
\nonumber\\
&& \;\;\;\; \times
\frac{1}{[\overset{L}{\underset{n=2}{\sum}}\bigl((k_{(n)})^{2}\bigr)^{n} 
+ x(t + \frac{u^{*}M^{2}}{2})]^{2- \frac{(d - \overset{L}{\underset{n=2}{\sum}}m_{n})}{2}}}\;\;\;.
\end{eqnarray}
Now we have to perform the remaining integral with 
higher power of momentum. Indeed, in the integral
\begin{equation}\label{10}
i_{n} =  \int \frac{d^{m_{n}}k_{(n)}}
{[\overset{L}{\underset{n=2}{\sum}}\bigl((k_{(n)})^{2}\bigr)^{n} + m^{2}]^{\gamma}},
\end{equation}
first perform the change of variables $r_{(n)}^{2}= 
k_{1(n)}^{2}+...+k_{m_{n}(n)}^{2}$. Second, change the variables to 
$z=r_{(n)}^{n}$ and then to $z'=z^{2}$. Collecting together these set of 
steps, we finally obtain 
\begin{equation}\label{11}
i_{n} = \frac{1}{2n \Gamma(\gamma)} S_{m_{n}}\Gamma(\frac{m_{n}}{2n}) 
\Gamma(\gamma - \frac{m_{n}}{2n}) (m^{2})^{-\gamma + \frac{m_{n}}{2n}}.
\end{equation}
Replacing this identity back into the expression for $I$, we can solve 
successfully all the integrals in higher powers of momentum along each 
competition subspace. Using $\epsilon_{L} = 4 + \overset{L}{\underset{n=2}{\sum}}\frac{(n-1)}{n} m_{n} - d$, we find
\begin{eqnarray}\label{12}
&& I = \frac{1}{2} S_{(d-\overset{L}{\underset{n=2}{\sum}} m_{n})} 
\Gamma(d - \overset{L}{\underset{n=2}{\sum}}m_{n}) \Gamma(2 -(d - \overset{L}{\underset{n=2}{\sum}}m_{n})) (\overset{L}{\underset{n=2}{\prod}} 
[\frac{S_{m_{n}} \Gamma(\frac{m_{n}}{2n})}{2n})])\nonumber\\
&& \;\;\times (t + \frac{u^{*}M^{2}}{2})^{-\frac{\epsilon}{2}}
(1+\frac{\epsilon}{2}+O(\epsilon^{2})),
\end{eqnarray}
which can be further simplified. In fact, developing the argument of the 
$\Gamma$-functions, using the identity 
$\Gamma(a + bx)= \Gamma(a)(1+bx \psi(a))$ and recalling to absorb the angular 
factor already mentioned above, it is not difficult to show that the integral 
has the following singular structure
\begin{equation}\label{13}
I= \frac{1}{\epsilon_{L}}(1+\epsilon_{L}[h_{m_{L}}-\frac{1}{2}(1 + 
ln(t+\frac{u^{*}M^{2}}{2}))]).
\end{equation}
We have to take another derivative of $H_{R}$ with respect to 
$M$, which will produce the inverse susceptibility
\begin{equation}\label{14}
\chi^{-1}= \frac{\partial H_{R}}{\partial M}= t + \frac{u^{*}M^{2}}{2} 
+ \frac{u^{*}}{4}(t + \frac{3u^{*}M^{2}}{2})[1+ln(t + \frac{u^{*}M^{2}}{2})] 
+ \frac{u^{* 2}M^{2}}{4}.
\end{equation}
For $T>T_{L}$ we substitute $M=0$ and the coupling constant at the fixed point 
value $u^{*}=\frac{2\epsilon_{L}}{3} + O(\epsilon_{L}^{2})$ into last equation, which produces the result
\begin{equation}\label{15}
\chi (T>T_{L}) = (1-\frac{\epsilon_{L}}{6})t^{-(1+\frac{\epsilon_{L}}{6})}.
\end{equation}   
Note that $\gamma_{L}= 1+\frac{\epsilon_{L}}{6}$ and $\chi(T>T_{L})$ above is 
consistent with scaling in the neighborhood of the critical point.

When $T<T_{L}$, we have to use the value of $M$ at the coexistence curve 
which is defined by the condition $H_{R}=0$, namely
\begin{equation}\label{16}
M^{2}= \frac{-6t}{u^{*}} + 3t[1 + ln(-2t)].
\end{equation} 
Replacing this value at expression (14) and neglecting 
$O(u^{* 2} \sim \epsilon_{L}^{2})$, we can demonstrate that the inverse 
susceptibility below the Lifshitz temperature has the form
\begin{equation}\label{17}
\chi^{-1}= (-2t)(1 + u^{*}(1 + ln[(-2t) + \frac{3u^{*}t}{2}(1+ln(-2t))])) 
+ \frac{3u^{*}t}{2}ln(-2t).
\end{equation}
Expanding the ``logarithm of the logarithm'' in the above expression using the 
expansion $ln(1+x)=x + O(x^{2})$, employing the fixed point 
$u^{*}=\frac{2\epsilon_{L}}{3} + O(\epsilon_{L}^{2})$ and neglecting 
$O(\epsilon_{L}^{2})$, we obtain
\begin{equation}\label{18}
\chi=(-t)^{-(1+\frac{\epsilon_{L}}{6})}\frac{1}{2}[1 - \frac{\epsilon_{L}}{6}
(4 +ln2)].
\end{equation}
Consequently, the susceptibility amplitude ratio is given by
\begin{equation}\label{19}
\frac{C_{+}}{C_{-}}= 2[1+\frac{\epsilon_{L}}{6}(3+ln2)]= 
2^{\gamma_{L} -1}\frac{\gamma_{L}}{\beta_{L}},
\end{equation}
where $\beta_{L}= \frac{1}{2} - \frac{\epsilon_{L}}{6}$. 
\par This expression is exact at one-loop level, its functional form 
in $\epsilon_{L}$ for several universal classes is the same, but the latter 
encodes {\it distinct} universalities since 
$\epsilon_{L}=\epsilon_{L}(d,m _{2},...,m_{L})$.

\section{Amplitude ratio for generic isotropic competing systems}
\par There are some minor modifications in the isotropic behaviors, but the 
trend to obtain the amplitude ratio follows the same script as in the 
anisotropic case. As there is only one subspace, say along 
$d= m_{n}$ space directions coupling $n$ neighbors via alternate competing 
interactions, the bare density Lagrangian is given by
\begin{eqnarray}\label{20}
L &=& \delta_{0n}  \frac{1}{2}
|\bigtriangledown_{m_{n}} \phi_0\,|^{2} 
+ \sum_{n'=2}^{n-1}\frac{1}{2} \tau_{nn'}
|\bigtriangledown_{m_{n}}^{n'} \phi_0\,|^{2} + \frac{\sigma_{n}}{2}
|\bigtriangledown_{m_{n}}^{n} \phi_0\,|^{2} + \nonumber\\
&&+ \frac{1}{2} t_{0}\phi_0^{2} + \frac{1}{4!}\lambda_0\phi_0^{4} .
\end{eqnarray}  
As before, the isotropic critical region is defined by 
$\delta_{0n}= \tau_{nn'}=0$ with $T \neq T_{L}$. There is only one 
renormalization group direction characterized by the $\xi_{n}$ 
correlation length. We perform a dimensional redefinition in the momentum just 
as done in the discussion of the anisotropic behavior. The expansion 
parameter is now $\epsilon_{n}=4n-m_{n}$. The renormalized free energy at 
one loop can be written in the form 
\begin{eqnarray}\label{21}
F(t,M)\;\; =&& \frac {1}{2}t M^{2} + \frac{1}{4!} g^{*} M^{4} +
\frac{1}{4}(t^{2} + g^{*} t M^{2} + \frac{1}{4} (g^{*} M^{2})^{2})
I_{SP} \nonumber\\
&& + \frac{1}{2} \int d^{m_{n}}k
\Bigl[ln\Bigl(1 + \frac{t+ \frac{1}{2} g^{*} M^{2}}{k^{2n}}\Bigr) \nonumber\\
&&- \frac{g^{*} M^{2}}{2 k^{2n}}\Bigr]\;\;,
\end{eqnarray}
and the nomenclature is almost the same as in the anisotropic case, except 
that now the integral $I_{SP}$ given by  
\begin{equation}\label{22}
I_{SP}= \Bigl[\int \frac{d^{m_{n}}k}{[(k+K')^{2n}]k^{2n}}\Bigr],
\end{equation}
is computed at the symmetry point $K^{' 2n}=\kappa^{2n}=1$. Performing a 
derivative with respect to $M$, we obtain  
\begin{equation}\label{23}
H_{R}=\frac{\partial F}{\partial M}= tM + \frac{1}{6} u^{*}M^{3} 
+ \frac{u^{*}M}{2}(t+ \frac{u^{*}M^{2}}{2})[I_{SP}-I],
\end{equation}
with 
\begin{equation}\label{24}
I =  \int \frac{d^{m_{n}}k}
{k^{2n} (k^{2n}+ t + \frac{u^{*}M^{2}}{2})}\;\;\;.
\end{equation}
\par Let us compute this last integral by employing a Feynman parameter, which 
gives essentially Eq.(\ref{7}) in the absence of the 
quadratic term. We then discover that the resulting integral has the same 
pattern as Eq.(\ref{11}) and can be solved along the same changes of variables 
in a identical manipulation which led to Eq.(\ref{12}). The geometric angular 
factor which appears here is just the area of the $m_{n}$-dimensional unity 
sphere $S_{m_{n}}$ and shall be absorbed in a redefinition of the coupling 
constant as before. Carrying out this procedure, we get to
\begin{equation}\label{25}
I =  \frac{1}{\epsilon_{n}}\Bigl[1-\frac{\epsilon_{n}}{2n}
ln\Bigl(t + \frac{u^{*}M^{2}}{2}\Bigr)\Bigr].
\end{equation}
\par Now, we can calculate the integral $I_{SP}$ either using the orthogonal 
approximation or exactly. Although we could determine the integral exactly, we 
would like to know the deviation between the two results. The reason is 
simple: in the computation of the exponents, only the anomalous dimension of 
the field had a significant difference: the first term is positive or negative 
depending the value of $n$ in the exact computation, but it is always positive 
in the orthogonal approximation. Nevertheless, a numerical analysis proved 
that the maximal error for increasing space dimension and (arbitraly) fixed 
$\epsilon_{n}=1$ occurred in the specific heat critical exponent 
for $n=2$ ($3.9\%$), increased for $n=3$ for the same exponent ($4.1\%$), but 
decreased for increasing $n$ ($n=4, 2.2\%$; $n=5, 3.4\%$, ...) \cite{LeiteII}. 
\par We have now, for the first time, the opportunity to test the 
effectiveness of the orthogonal approximation in amplitudes, which in our 
opinion is worthy analyzing in view of the facts already known from the 
deviations of critical exponents. 
\par Next, let us perform the computation of the amplitude ratio using either 
the orthogonal approximation or the exact computation of the integral 
$I_{SP}$. As we shall see in a moment, the amplitudes above and below the 
Lifshitz temperature change.

\subsection{Amplitude ratio using the orthogonal approximation}
According to Ref. \cite{LeiteII}, the integral computed at the symmetry point 
utilizing the orthogonal approximation was shown to result in the expression 
$I_{SP}= \frac{1}{\epsilon_{n}}(1+\frac{\epsilon_{n}}{2n})$. Hence,
\begin{equation}\label{26}
I_{SP}-I= \frac{1}{2n}\Bigl[1+ln\Bigl(t + \frac{u^{*}M^{2}}{2}\Bigr)\Bigr],
\end{equation}
which implies
\begin{equation}\label{27}
H_{R}=\frac{\partial F}{\partial M}= tM + \frac{1}{6} u^{*}M^{3} 
+ \frac{u^{*}M}{4n}(t+ \frac{u^{*}M^{2}}{2})\Bigl[1+ln\Bigl(t + \frac{u^{*}M^{2}}{2}\Bigr)\Bigr].
\end{equation}
Taking another derivative with respect to $M$, we find
\begin{equation}\label{28}
\chi^{-1}= t + \frac{u^{*}M^{2}}{2} 
+ \frac{u^{*}}{4n}(t + \frac{3u^{*}M^{2}}{2})[1+ln(t + \frac{u^{*}M^{2}}{2})] 
+ \frac{u^{* 2}M^{2}}{4n}.
\end{equation}
Now, $M=0$ in last equation is the situation corresponding to $T>T_{L} (t>0)$, 
or in other words
\begin{equation}\label{29}
\chi^{-1}(T>T_{L})= t + \frac{u^{*} t}{4n} [1+lnt].
\end{equation} 
Replacing the fixed point value $u^{*}=\frac{2\epsilon_{n}}{3}$, the 
susceptibility above $T_{L}$ reads
\begin{equation}\label{30}
\chi(T>T_{L})= t^{-\gamma_{n}}\Bigl(1-\frac{\epsilon_{n}}{6n}\Bigr),
\end{equation}
where $\gamma_{n}=1+\frac{\epsilon_{n}}{6n}$ is the isotropic susceptibility 
exponent. Below $T_{L}$, we determine the value of $M$ in the coexistence 
curve defined by $H_{R}=0$, which yields
\begin{equation}\label{31}
M^{2}= \frac{(-6t)}{u^{*}} + \frac{3t}{n}[1+ln(-2t)].
\end{equation}
Replacing this value in the expression of $\chi^{-1}$, it leads to
\begin{equation}\label{32}
\chi^{-1}(T<T_{L}) = (-2t)[1 + \frac{u^{*}}{4n}ln(-2t) +  \frac{u^{*}}{n}],
\end{equation}
which at the fixed point $u^{*}=\frac{2}{3\epsilon_{n}}$ implies that we 
can write the susceptibility in the form
\begin{equation}\label{33}
\chi(T<T_{L})= (-t)^{\gamma_{n}} \frac{1}{2}[1-\frac{\epsilon_{n}}{6n}(4+ln2)].
\end{equation}
The susceptibility amplitude ratio which results from the above expressions 
is written as 
\begin{equation}\label{34}
\frac{C_{+}}{C_{-}}= 2\Bigl[1 + \frac{\epsilon_{n}}{2n} 
+ \frac{\epsilon_{n}}{6n} ln2 \Bigr]= 
2^{\gamma_{n}-1}\frac{\gamma_{n}}{\beta_{n}},
\end{equation} 
where $\beta_{n}= \frac{1}{2} - \frac{\epsilon_{L}}{3n}$ is the magnetization 
exponent.
\subsection{Exact amplitude ratio}
The main advantage of the isotropic case is that the Feynman integrals can be 
computed exactly. Thus, we can obtain the susceptibility amplitude ratio 
without the necessity of using approximations.  The $I_{SP}$ integral was 
already computed in Ref. \cite{LeiteII} at the symmetry point and was shown 
to be given by the expression 
$I_{SP}= \frac{1}{\epsilon_{n}}[1 + D(n)\epsilon_{n}]$, where 
$D(n) = \frac{1}{2}[\psi(2n) + \psi(1)] - \psi(n)$. First, using Eq. (25) 
we find
\begin{equation}\label{35}
I_{SP}-I= D(n) + \frac{1}{2n} ln\Bigl(t + \frac{u^{*}M^{2}}{2}\Bigr),
\end{equation}
which turns out to result in the following magnetic field
\begin{equation}\label{36}
H_{R}=\frac{\partial F}{\partial M}= tM + \frac{1}{6} u^{*}M^{3} 
+ \frac{u^{*}M}{4n}(t+ \frac{u^{*}M^{2}}{2})\Bigl[2n D(n)+ln\Bigl(t + \frac{u^{*}M^{2}}{2}\Bigr)\Bigr].
\end{equation}
It is easy to show that the inverse susceptibility which follows can be 
written as
\begin{equation}\label{37}
\chi^{-1}= t + \frac{u^{*}M^{2}}{2} 
+ \frac{u^{*}}{4n}(t + \frac{3u^{*}M^{2}}{2})[2n D(n)+ln(t + \frac{u^{*}M^{2}}{2})] 
+ \frac{u^{* 2}M^{2}}{4n}.
\end{equation}
Hereafter we are going to use the coupling constant at the fixed point, i.e., 
$u^{*}=\frac{2\epsilon_{n}}{3}$. Set $M=0$ for $T>T_{L}$ in the above equation 
in order to find the susceptibility in the following form
\begin{equation}\label{38}
\chi(T>T_{L})= t^{-\gamma_{n}}\Bigl(1-\frac{D(n)\epsilon_{n}}{3}\Bigr).
\end{equation}
For $T<T_{L}$ the value of $M$ in the coexistence curve is given by
\begin{equation}\label{39}
M^{2}= \frac{(-6t)}{u^{*}} + \frac{3t}{n}[2n D(n)+ln(-2t)].
\end{equation}
Substitution of this value into the expression for $\chi^{-1}$ results in 
the following value for the susceptibility below $T_{L}$
\begin{equation}\label{40}
\chi(T>T_{L})= (-2t)^{-\gamma_{n}}
\Bigl[1-\frac{\epsilon_{n}}{2n}\Bigl(1+ \frac{2nD(n)}{3}\Bigr) \Bigr].
\end{equation}   
Using Eqs. (\ref{38}) and (\ref{40}), we finally obtain
\begin{equation}\label{41}
\frac{C_{+}}{C_{-}}= 2\Bigl[1 + \frac{\epsilon_{n}}{2n} 
+ \frac{\epsilon_{n}}{6n} ln2 \Bigr]= 
2^{\gamma_{n}-1}\frac{\gamma_{n}}{\beta_{n}},
\end{equation} 
which is the same value obtained using the orthogonal approximation 
Eq. (\ref{34}). 
As happened to the critical exponents at one-loop order, the orthogonal 
approximation and the exact computations of the susceptibility amplitude ratio 
in the isotropic case yield the same value. However, we do expect deviations 
in both calculations at two-loop order and beyond.
\section{Conclusion}  
\par The obtained anisotropic amplitude ratios maintain the same functional 
form as its counterpart in the Ising-like universality class, with the 
parameter $\epsilon_{L}= = 4 + \overset{L}{\underset{n=2}{\sum}}\frac{(n-1)}{n} m_{n} - d$ replacing the ordinary perturbation parameter of noncompeting 
systems $\epsilon=4-d$. This leads to the property of universality class 
reduction. This property already 
appeared in the computation of exponents and is expected to be valid at all 
loop orders.
\par In fact, if we turn off all the competing interactions, i. e., by setting 
$m_{3}= ...=m_{L}=0$, keeping just 
alternate couplings among second neighbors and identify $m_{2}=m$, we obtain 
the result for the anisotropic $m$-axial universality class. Note that the 
ratio has the same functional form for all values of $m \neq d$ and reproduces 
the uniaxial case $m=1$ studied earlier \cite{Leite3}. For instance, 
three-dimensional systems have perturbative parameters 
$\epsilon_{L}= 1 + \frac{m}{2}$ which change at distinct values of $m$. 
Consequently, they produce different values for the amplitude ratios which is 
consistent with the universality hypothesis previously stated. Besides, if go 
on and switch off the competing interactions among second neighbors ($m=0$) we 
obtain the result of the Ising-like universality class. 
\par The isotropic amplitude ratios, on the other hand, possesses its own 
version of universality class reduction. Different values of the number of 
neighbors coupled via alternate couplings ($n$) are responsible by the 
variation of the susceptibility ratio. The case $n=2$ corresponds to the 
isotropic $m$-axial ($d=m$ close to 8) universality class. Our result is 
the first computation of isotropic amplitude ratios, perhaps because 
isotropic systems were found only in lower dimensional systems ($m=d=3$) 
so far, as proposed in the critical behavior of homopolymer-diblock 
copolymer mixtures \cite{BF,KM,Pipi,Natalie}, which makes the perturbative 
parameter rather large ($\epsilon_{2}=5$). Let us try to extract meaningful 
results from our results for those three-dimensional systems.
\par Although the value of the amplitude above the critical temperature is not 
universal, let us compare the two values using the approximate and exact 
results for the isotropic case $n=2$. Using $\epsilon_{2}=5$ in Eq.(\ref{30}) 
we find $C_{+}=0.583$ using the orthogonal approximation, whereas the exact 
computation from Eq.(\ref{38}) using $D(2)=-\frac{1}{12}$ yields 
$C_{+}=1.13$, and the deviation is huge. Nevertheless, comparing with the 
Table III from Ref. \cite{Pipi} both values are allowed. In fact for 
diblock polymer composition $\Phi_{DB}=0.072$ at temperature 
$69.2 \pm 0.1 ^{o}C$ the measured amplitude is given by $C_{+}=0.6 \pm 0.04$ 
which is compatible with the value obtained via the orthogonal approximation. 
On the other hand, for a slightly change of composition, namely 
$\Phi_{DB}=0.073$ measured  at temperature $69.5 \pm 0.2 ^{o}C$ the amplitude 
value is $C_{+}=0.94 \pm 0.07$, which is also consistent with the exact 
amplitude. Notice that even though those authors confirmed the absence of 
the isotropic Lifshitz point, they considered the Lifshitz critical region 
{\it with the inclusion of the fluctuations} using SCFT (Ref. \cite{KM}) in 
their experimental fits of the susceptibility curves, which is quite 
a different method than the one proposed in the present work \cite{Smirnova}. 
\par Therefore, this is the first solid indication that field theory 
renormalization group results including the contribution of fluctuations are 
consistent with experiments in those sort of polymers, in spite of the large 
value of the perturbative parameter. Though the deviations between the 
amplitudes are significant and expected from their nonuniversal feature, 
the experimental results do not rule out the orthogonal approximation result. 
This is the first experimental ground to test the deviations in both 
calculations. But we can go on and compare the true universal susceptibility 
exponent obtained in the experiment with our previous two-loop calculation 
from Ref.\cite{LeiteII}. The orthogonal approximation for $(N=1, d=m=3)$ 
yields $\gamma_{2}=1.90$, whereas the exact exponent is $\gamma_{2}=1.50$. The 
latter is consistent with the experimental value $\gamma_{2}=1.55 \pm 0.15$ 
obtained from the isotherm at 69.5$^{o}$C with concentration of diblock 
copolymer at $\Phi_{DB}=0.071$. It is amazing that the experiment carried out 
on the homopolymer-diblock copolymer considered by those authors can really be 
described using the isotropic Lifshitz universality class and its critical 
region, in spite of the large value of the perturbative parameter for 
three-dimensional systems. Perhaps the use of other field-theoretic isotropic 
results already (and to be) developed in other experiments to be performed 
might be successful in refining our knowledge of the Lifshitz critical region 
for these systems.
\par Another aspect is the theoretical possibility of occurrence of up to 
$6th$ character Lifshitz points in $AB/BC$ mixtures of diblock copolymers 
\cite{OH}. If this system can be fabricated in the laboratory, our work 
represents a prevision of results for the susceptibility with increasing values of the perturbation parameter for three-dimensional systems, in analogy to 
what was studied in Ref. \cite{Pipi} using small angle neutron scattering. 
This is rather encouraging an evidence to pursue further 
universal aspects of this kind of critical behaviors using this field 
theoretical language, for instance, amplitude ratios above and below the 
critical temperature. This could shed new light in devising experimental 
applications to our model in order to measure those effects in a real 
physical system, with isotropic or (less obvious) anisotropic critical 
behaviors.
\par The universality class reduction in the isotropic case is even more 
evident than its anisotropic counterpart. As a matter of fact, $n=1$ 
corresponds the system without competition 
and belongs to the 
Ising-like universality class. Therefore, systems without competition can be 
understood as special cases either from the anisotropic cases 
($m_{n}=0, n=2,...,L$) or from the isotropic cases $n=1$, a property already 
discovered in the computation of the critical exponents.     
\par Since the isotropic ratio can be computed approximately and exactly as 
well, we calculated the ratio using both procedures for the sake of comparison 
of the deviations for individual amplitudes and how this deviation could be 
understood at least in the case $n=2$. The amplitudes themselves are different 
in both cases, but the ratio is equal. This property also takes place in the 
determination of critical indices using perturbation theory, but the result is 
valid only at one-loop level. We expect that both ratios will have deviations 
at two-loop order.
\par The most interesting extension of the method proposed here is the study 
of the specific heat amplitude ratio for generic competing systems, 
generalizing the discussion carried out for the anisotropic $m$-axial 
critical behavior\cite{Leite3}. It would be nice to tackle the computation 
of other universal amplitude ratios either at one-loop level or to extend the 
method to compute amplitude ratios at two-loop order \cite{Ber} for generic 
competing systems.
\par Last but not least, we hope that the present investigation can be 
significant to motivate experimental techniques in order to determine  
the susceptibility amplitude ratio in magnetic systems such as $MnP$, 
$Mn_{0.9}Co_{0.1}P$ \cite{MnCoP,Pla}, etc.. In addition the new phase 
encountered in $MnP$ \cite{new} and $Mn_{0.9}Co_{0.1}P$ \cite{Pla} 
might be related to new effects of competition as described in the present 
work.

\section{Acknowledgments}
CFF would like to thank CAPES for financial support.

\end{document}